\documentclass[%
 reprint,
superscriptaddress,
 amsmath,amssymb,
 aps,
 prl
]{revtex4-2}

\usepackage{graphicx}
\usepackage{dcolumn}
\usepackage{bm}

\usepackage{color}
\usepackage{amssymb}
\usepackage[unicode=true,pdfusetitle,
 bookmarks=true,bookmarksnumbered=true,bookmarksopen=true,
 breaklinks=true,pdfborder={0 0 1},colorlinks=true]
{hyperref}
\hypersetup{linkcolor=blue,anchorcolor=blue,citecolor=blue,urlcolor=blue}


\begin{document}

\preprint{APS/123-QED}

\title{Ghost Panorama}

\author{Zhiyuan Ye}
\author{Hai-Bo Wang}
\author{Jun Xiong}
 \email{junxiong@bnu.edu.cn}
\author{Kaige Wang}

\affiliation{Department of Physics, Applied Optics Beijing Area Major Laboratory, Beijing Normal University, Beijing 100875, China}

\begin{abstract}

Computational ghost imaging or single-pixel imaging enables the image formation of an unknown scene using a lens-free photodetector. In this Letter, we present a computational panoramic ghost imaging system that can achieve the full-color panorama using a single-pixel photodetector, where a convex mirror performs the optical transformation of the engineered Hadamard-based circular illumination pattern from unidirectionally to omnidirectionally. To our best knowledge, it is the first time to propose the concept of \textit{ghost panorama} and realize preliminary experimentations. It is foreseeable that \textit{ghost panorama} will have more advantages in imaging and detection in many extreme conditions (e.g., scattering/turbulence, cryogenic temperatures, and unconventional spectra), as well as broad application prospects in the positioning of fast-moving targets and situation awareness for autonomous vehicles. 

\end{abstract}


\maketitle

The idea of using a photodiode to perceive the world has long been a reality since Shih and his co-workers realized the first ghost imaging (GI) \cite{Pittman1995} in 1995. Whether through quantum or classical optical correlations\cite{Caodz2005,Moreau2018}, or through a spatial light modulator computationally\cite{Duarte2008,Shapiro2008}, as long as the light field illuminating onto the object can be directly recorded or indirectly accessed\cite{Paniagua-Diaz2009,Yez2020}, the image of the object can be fully retrieved, even through the extreme environment, such as turbulence\cite{Chengj2009,MeyersRE2011}, scattering\cite{Escobet2018} and cryogenic temperatures\cite{MedaA2015}. Also, GI modality provides a low-cost\cite{Edgar2019} and promising solution\cite{YuH2016,ZhangAX2018,Chanwl2008,OlivieriL2018} in some spectra where image sensors are expensive or even nonexistent. These novel properties and prospects have attracted a lot of attention such that GI has been widely applied in 3D imaging\cite{SunB2013,SunMJ2019}, multi-spectral imaging\cite{WelshSS2013,BianL2016}, light field imaging\cite{Angelo2016,Peng2018}, leak gas imaging\cite{Gibson2017}, cytometry\cite{Ota2018}, tomography\cite{Kingston2018}, etc. 

To date, however, panoramic GI has not been discussed or reported, and it seems a fantasy to use a single photodiode to obtain an omnidirectional pixelated image. Conventional panoramic imaging systems can reach 360 degrees in the horizontal direction, which are widely used in the fields of robot vision, surveillance, video conferencing and even external space detection. In general, traditional panoramic imaging has two paths: One is to digitally process multiple images from various angles via image stitching and fusion\cite{Brown2007}, which has been fully developed, just like 3D imaging, such that one can even use a smartphone to generate a panorama; Another solution is to adopt a special lens group (e.g., fisheye lens)\cite{Powell1994,Kweon2005,Pernechele2016} and reflective surface\cite{Chahl1997,Baker1999,JangG2006,LiW2011} to project the omnidirectional scene onto the target surface of the image sensor, and then digitally process it to obtain a panorama. Although the former is user-friendly and popular without the need of special lens or reflective surfaces, the latter has higher imaging efficiency and lower computational complexity. At this stage, almost all panoramic imaging technologies rely on lens-based image sensors to become a convention, which makes people ignore some facts: in some non-visible bands, conventional lenses are more difficult to manufacture and optimize compared to reflective surfaces, and the cost of high-resolution image sensors is expensive. These inherent limitations prevent the application of panoramic imaging in a wider range of conditions. In this Letter, we will present a new path to achieve panoramic imaging using one single-pixel detector and a convex mirror, named as \textit{ghost panorama}, breaking through the common sense and limitations to panoramic imaging.

Before extending to panoramic imaging, we review the principle of computational GI (CGI) modality. To sample an object $O$, CGI generally requires a series of known lighting patterns $\left \{P\right \}$ to illuminate the object and a photodetector to measure the backscattered intensity $\left \{I\right \}$, with a second-order intensity correlation function $G^{(2)}$ to reconstruct the ghost image:
\begin{equation}
G^{(2)}=\left \langle PI\right \rangle-\left \langle P\right \rangle\left \langle I\right \rangle\propto O,
\label{eq:1}
\end{equation}
where $\left \langle \right \rangle$ denotes ensemble average. It appears that the family of the illumination mode $\left \{P\right \}$ plays a very critical role in CGI, directly determining the imaging quality and efficiency. Common illumination modes include Hadamard basis\cite{Pratt1969,Wangl2016}, Fourier basis\cite{Zhangz2015,Zhangz2017}, etc. In almost all CGI systems, the propagation direction of the light field is unidirectional, which is considered as a common sense. Hence, how to achieve omnidirectional and accessible (known) structured illumination is the key to achieving \textit{ghost panorama}. 

\begin{figure*}[!ht]
\centering
{\includegraphics[width=15cm]{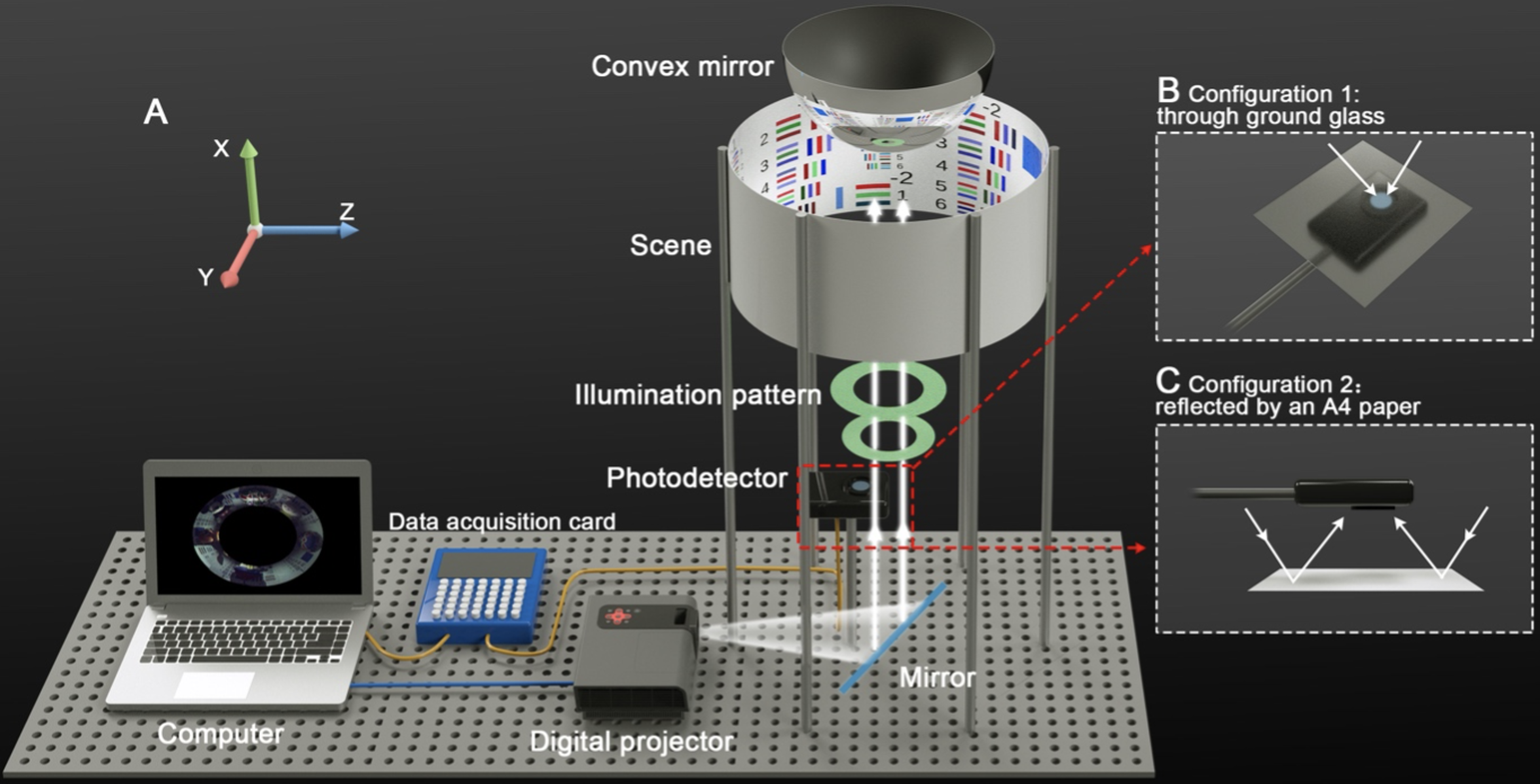}}
\caption{Schematic experimental setup (A) of \textit{ghost panorama}: a color commercial projector (Sony, VPL-EW578) functions as a time-varying light source to play lighting patterns onto the scene sequentially, while a horizontally lying single-pixel photodetector (Thorlabs, PDA100A2) free of lens continuously collects the backscattered light intensity synchronously. A shielded junction box (NI, BNC-2110) and a data acquisition card (NI, PCIe-6251) are used to record the signal, and a host computer processes all the raw data. The sampling rate of the data acquisition card is set to 60KS/s. A convex mirror (curvature radius is 30cm) completes an optical transformation of the patterns propagating from unidirectionally to omnidirectionally. The photodetector receives the scattered light in two manner: through ground glass (B); reflected through A4 white paper (C).}
\label{fig1}
\end{figure*}
Inspired by the conventional catadioptric panoramic imaging systems\cite{Chahl1997}, we utilize a convex mirror as an optical transformation to generate omnidirectional illumination shown in Fig. \ref{fig1}(A). Specifically, the convex mirror can expand an illumination pattern propagating in a single direction into an omnidirectional illumination field along the cylinder. Of course, the convex mirror will bring inherent distortion to the lighting patterns, so we should only select the appropriate region of the convex mirror. Besides, the effective number of pixels in illumination patterns determine the resolution of the panorama, and accordingly determine sampling time, and computational complexity as well.

\begin{figure}[!h]
\centering
{\includegraphics[width=\linewidth]{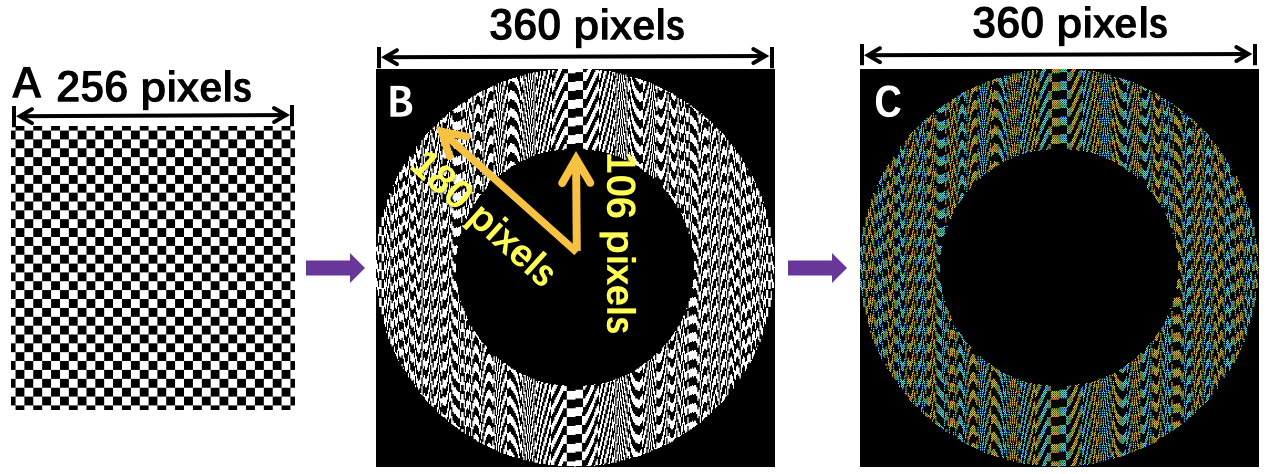}}
\caption{Generation of a circular illumination pattern. (A) An example of Hadamard-based lighting patterns. (B) Hadamard-based circular illumination pattern by linear mapping of (A). (C) An example of Hadamard-Bayer circular illumination patterns. The radius of the small circle is about 106 pixels and the radius of the large circle is 180 pixels. One can customize the size of the ring spacing to match the convex mirror.}
\label{fig2}
\end{figure}

In order to balance these constraints well and match the size of the convex mirror, we carefully design a circular illumination pattern (Fig. \ref{fig2}(B)) based on the original Hadamard basis (Fig. \ref{fig2}(A)) using a pixel-to-pixel linear mapping scheme\cite{Yez2019}. The design of the ring-shaped illumination patterns can effectively select the imaging area of interest. By adjusting the relative size of the patterns and the convex mirror, an appropriate region with less distortion can be selected. We expect the size of reconstructed image to be $256\times256$ pixels and the set ring spacing to be approximately equal to 74 pixels. In order to quickly achieve a full-color panorama, we apply a Bayer array\cite{Bayer1976} on the original circular pattern (Fig. \ref{fig2}(B)) to obtain a Hadamard-Bayer circular pattern (Fig. \ref{fig2}(C))\cite{Yezzy2020}, which eliminates the need for multiple detectors carrying color filters or multiple repeated measurements for different colors. The Bayer array simulates the sensitivity of the human eye to color, and uses a certain arrangement (for example, red filter cells : green filter cells : blue filter cells = 1:2:1) to convert grayscale information into color information. In other words, we can achieve a full-color ghost panorama with a single detector at a sampling rate of $33.33\%$ naturally, only requiring a Bayer-array-based interpolation algorithm\cite{Gunturk2005} to demosaic the reconstructed gray panorama.

\begin{figure*}[htbp]
\centering
{\includegraphics[width=16cm]{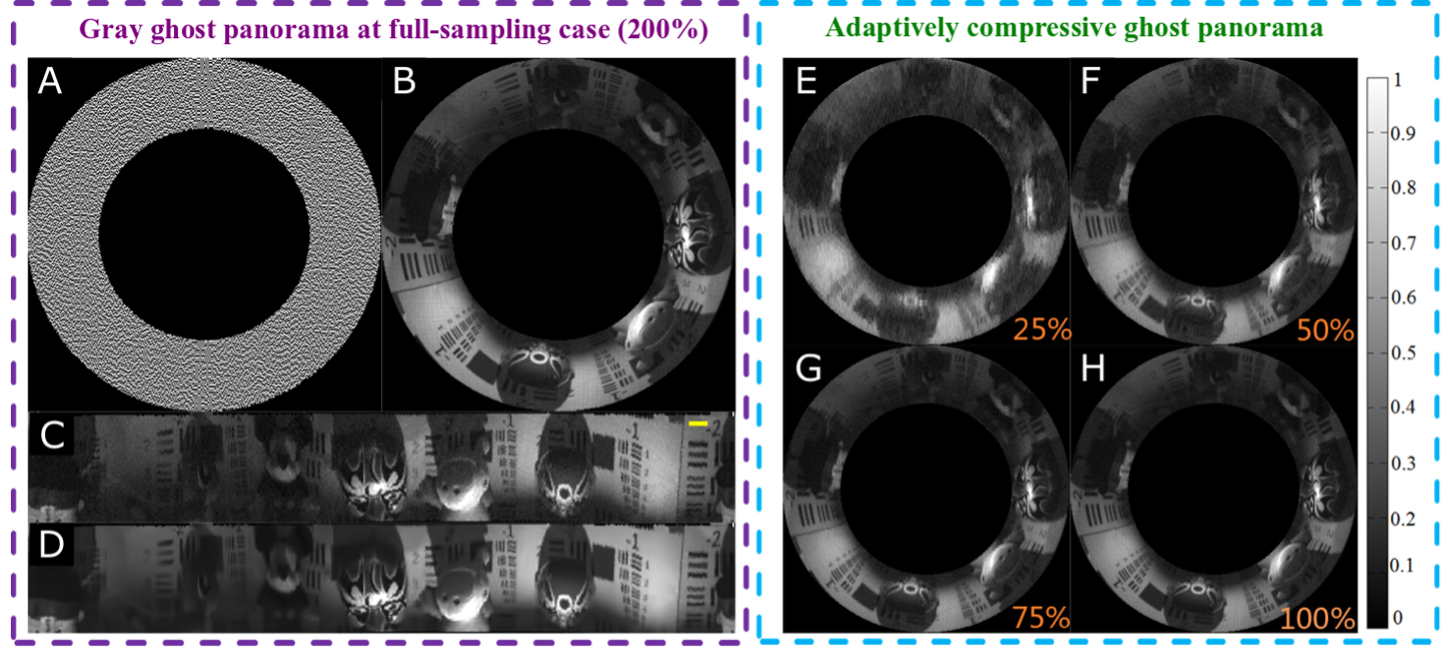}}
\caption{Experimental results of gray ghost panoramas. (A) An example of gray circular illumination patterns. (B) The gray ghost panorama at full-sampling condition (the sampling rate is $200\%$). (C) The unrolled omnidirectional ghost image. (D) The unrolled omnidirectional ghost image using a BM3D filter. (E to H) Adaptively compressive ghost panoramas at sampling rate of $25\%$, $50\%$, $75\%$, $100\%$, respectively. Scale bar = 4 cm.}
\label{fig3}
\end{figure*}

In a proof-of-principle experiment, we choose a color commercial projector as the structured light source. Under the full-sampling condition, we need to play $256\times256\times2$ ring engineered patterns (Fig. \ref{fig3}(A)) to sample an omnidirectional object. A flat-lying photodetector captures the temporal intensity signal from the backscattered light simultaneously. Figs. \ref{fig3}(B) and (C) show the gray ghost panorama and the unrolled omnidirectional ghost image of the scene through a circular expansion algorithm\cite{Chahl1997}, respectively.  A BM3D filter\cite{Danielyan2009} is utilized to further enhance the quality of the original ghost panorama shown in Fig. \ref{fig3}(D) (this procedure is optional). Due to the limitations of resolution, the field of view of \textit{ghost panorama} is about 360 degrees (horizontal) by 60 degrees (longitudinal) at this stage, which is determined by a trade-off between the number of effective pixels and the size of the engineered ring pattern. To effectively alleviate a large amount of sampling time, we now demonstrate an adaptively compressive sampling strategy\cite{Zhangz2017} based on the nature of Hadamard basis itself. Similar to the Fourier transform, the projection of natural images into the Hadamard domain\cite{Pratt1969} is generally sparse as well, that is, a small number of coefficients have larger weights, characterizing the main features of the image. So, we can use this feature to achieve adaptive image reconstruction without the need for an extra time-consuming optimization algorithm. Then all we need to do is to play the lighting patterns in a proper order\cite{Vaz2020}, in which the information corresponding to the ``low frequency area” will be collected preferentially as a preview of the scene, and the ``high frequency area” will be subsequently acquired to repair the details. However, since the circular Hadamard-based illumination pattern undergoes a random linear mapping of the original mode, the order requires some additional corrections and optimizations (See \textbf{Supplement} Section \textbf{S1}). Figs. \ref{fig3}(B) and (E) to (H) show the experimental results at different sampling rate, $200\%$ (B), $25\%$ (E), $50\%$ (F), $75\%$ (G), $100\%$ (H), respectively. \textbf{Visualization1} demonstrates this dynamic sampling process using four different orders, and the performance analysis of the optimized compressive sampling strategy is shown in Section \textbf{S1} of \textbf{Supplement}.

\begin{figure*}[!ht]
\centering
{\includegraphics[width=16cm]{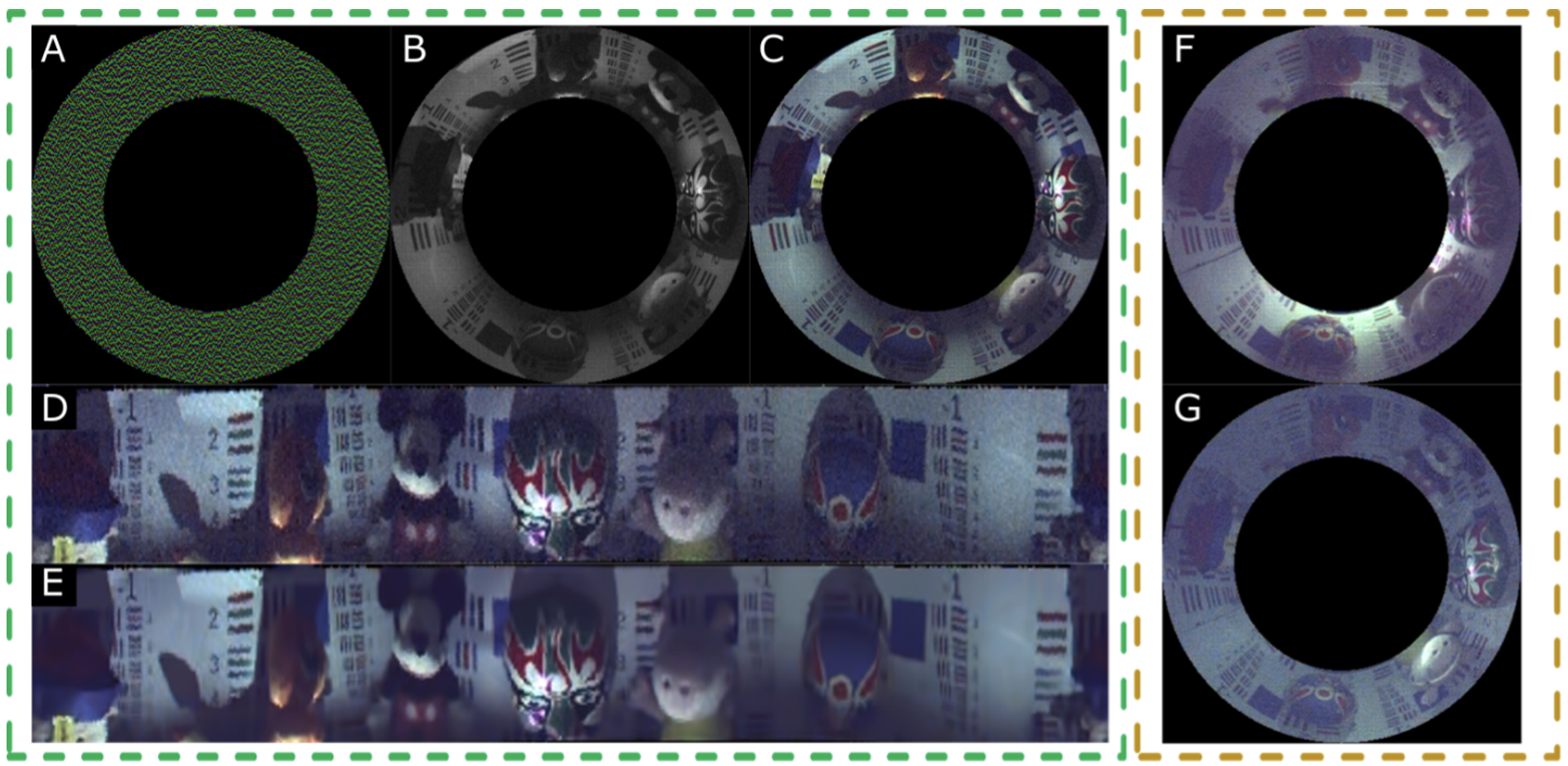}}
\caption{Experimental results of full-color ghost panoramas. (A) An example of circular illumination patterns colored by the Bayer array. (B) The reconstructed mosaic ghost image. (C) The demosaicing image of full-color ghost panorama. (D) The unrolled full-color omnidirectional ghost image. (E) The unrolled full-color omnidirectional ghost image using a BM3D filter. Full-color ghost panoramas in the noisy environment: (F) through thick ground glass and (G) reflected by A4 white paper.}
\label{fig4}
\end{figure*}

As for full-color \textit{ghost panorama}, $256\times256\times2$ Hadamard-Bayer patterns (Fig. \ref{fig4}(A)) are projected onto the scene at a sampling rate of $33.33\%$. Figure \ref{fig4}(B) shows the directly reconstructed ghost panorama, and Figures \ref{fig4}(C) to (E) show the experimental results of full-color ghost panoramas using the image interpolation algorithm. Next we demonstrate another inherent feature of GI modality, that is, we can achieve panoramic imaging in the presence of noisy environments. As shown in Figs. \ref{fig1}(B) and (C), we add a piece of thick ground glass (transmissive light) and a piece of A4 white paper (reflective light) in front of the receiving end, and repeat this experiment twice. As shown in Figs. \ref{fig4}(F) and (G), the full-color ghost panoramas can still be retrieved with the inevitable reduction of the signal-to-noise ratio due to the weakening of the light intensity signal. This result means that ghost panoramas can be transmitted freely in noisy environments such as strong scattering. In fact, in more extreme environments, such as low light and cryogenic temperatures, only one photodiode in the receiving end without any optical support element will bring unprecedented flexibility and achieve unexpected image reconstruction with higher image quality compared to conventional imaging technologies. 

It is worth noting that the lens-free photodetector must be placed flat in a suitable position to collect scattered light from all directions as much as possible. It is equally low-cost and feasible to use flexible photosensitive materials to develop curved photodetection targets. \textit{Ghost panorama} is expected to have a broader application prospect and development potential: If the projection-lens-assisted structured illumination is replaced with the pseudo-thermal light\cite{Caodz2005}, lensless \textit{ghost panorama} can be achieved (See Section \textbf{S2} for the schematic diagram in the \textbf{Supplement}); if multiple photodetectors or one photodetector with time resolution is used, 3D \textit{ghost panorama} might be accessible; if the intensity-coded illumination\cite{DengQ2020} is applied, non-imaging tracking omnidirectionally of a fast-moving object could be realized. These novel frontiers will greatly promote the development of GI, making it better applied in the real world, such as omnidirectional 3D situation awareness for autonomous vehicles\cite{Edgar2019}. 

To conclude, full-color panoramic imaging through scattering media, with a lens-free single-pixel photodetector for the first time, is demonstrated preliminarily. At last, to end up with another fantasy: Is it possible to generate a known or accessible fluctuating light field, like black body radiation or the sun, that emits a full-space light field into all directions, and a spherical detector (similar to an integrating sphere) collect scattered light from all directions, so as to achieve full-space GI eventually.

See \textbf{Supplement} for supporting the content.

\end{document}